# CONDUCTING VERIFICATION AND VALIDATION OF MULTI- AGENT SYSTEMS

A. Al-Neaimi[1], S. Qatawneh[2], Nedhal Al Saiyd [3]

[1]Department of Software Engineering, AL-Zaytoonah University, Amman, Jordan
[2]Department of Software Engineering, AL-Zaytoonah University, Amman, Jordan
[3]Department of Computer Science, Applied Science University, Jordan

## ABSTRACT

*Verification and Validation (V&V) is a series of activities, technical and managerial, which performed by system tester not the system developer in order to improve the system quality, system reliability and assure that product satisfies the users operational needs. Verification is the assurance that the products of a particular development phase are consistent with the requirements of that phase and preceding phase(s), while validation is the assurance that the final product meets system requirements. an outside agency can be used to performed V&V, which is indicate by Independent V&V, or IV&V, or by a group within the organization but not the developer, referred to as Internal V&V. Use of V&V often accompanies testing, can improve quality assurance, and can reduce risk. This paper putting guidelines for performing V&V of Multi-Agent Systems (MAS).*

## KEYWORDS

*verification and validation life cycle , Multi-agent system, verification and validation Techniques*

## 1. INTRODUCTION

Over the past three decades, software engineers have derived a progressively better understanding of the characteristics and complexity of designing and implementing agent-based applications[1]. IA (Intelligent Agent) can be defined as an entity that has a certain amount of intelligence and can perform a set of tasks on the behalf of users, due to its autonomy, mobility, and reactivity[2].

MAS (Multi-Agent Systems) are used as a paradigm for conceptualizing, designing, and implementing software engineering systems that are beyond the individual knowledge of the traditional centralized approaches [3]. MAS have its own problem solving capabilities and the agents are able to interact to reach goals across flexible and distributed networks of information sources and expertise. In order to assuring V&V (verification and validation)(Figure 1)[4] in developing MAS it need to involves the measurement and assessment of a variety of quality characteristics such as correctness, usability, reusability, competency and consistency [5]. MAS Verification deals with the building of correct system. This process is substantiating that the system is transformed from one into another, as intended, with sufficient accuracy. MAS Verification is a two-stage process. The first process is the transformation of a formulation problem into a specification. The second process is the conversion of system representation form into an executable computer program. During the V&V process, MAS is subjected to a several testing forms to perform either verification or validation or V&V. Some tests (verification) are

International Journal of Software Engineering & Applications (IJSEA), Vol.3, No.5, September 2012

intended to judge the accuracy of system transformation from one form into another, and some tests (validation) are devised to evaluate the behavioral accuracy of the system[6].

Our paper is to present guidelines for conducting V&V of MAS. Section 2 recalls the basic elements related to V&V .Section 3 discusses the V&V activities in the MAS life cycle. Section 4 discusses the V&V techniques. Section 5 gives conclusion and describes our future work.

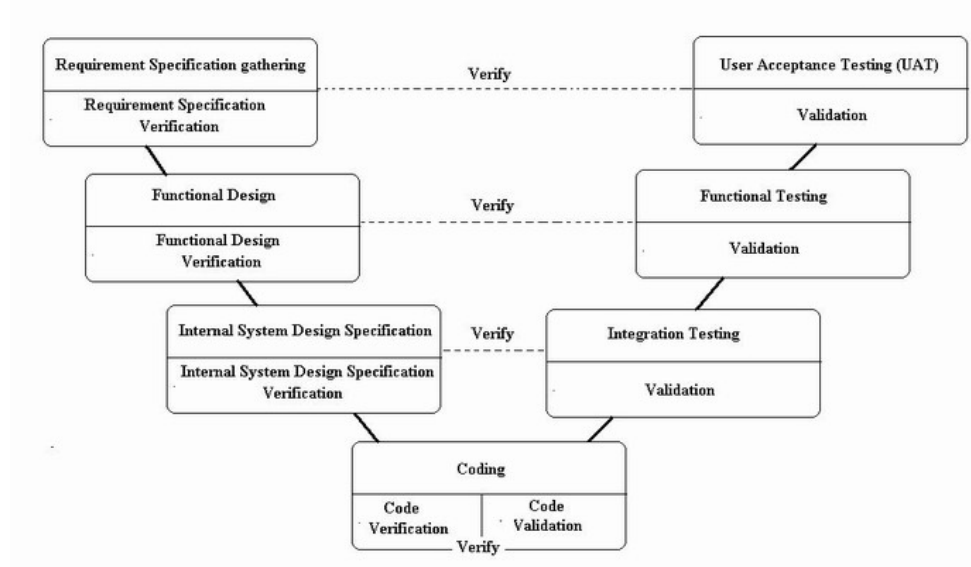

Figure 1.verification & validation model

## 2. BACKGROUND AND RELATED WORKS

### 2.1. MAS

MAS have emerged as one of the most important areas of research and development in information technology. A multi-agent system composed of multiple interacting software components known as agents, which are typically capable of cooperating to solve problems that are beyond the abilities of any individual member[7]. Multi-agent systems are important primarily because they have been found to have very wide applicability, in areas as diverse as industrial process control and electronic commerce[2](Figure 2 is an example of MAS).

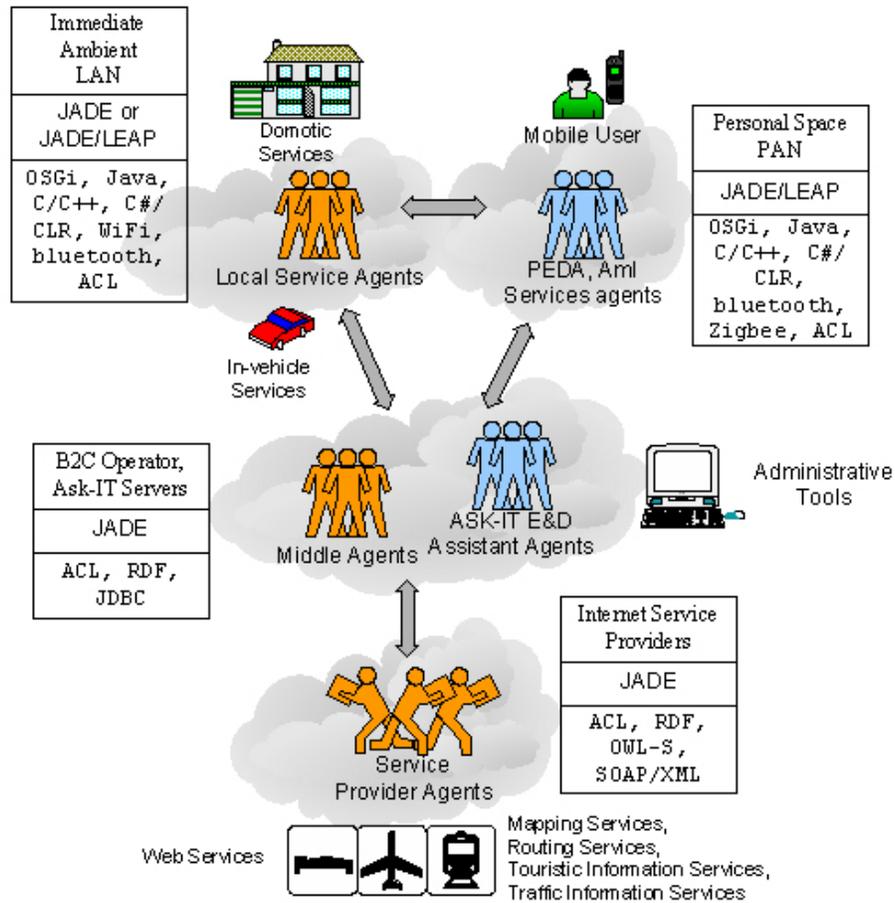

Figure 2. Multi _Agent System Example

## 2.2. V&V Attitudes

V & V attitudes are important to understand the foundations of V&V process[8]. The attitudes help the researchers in understanding the V&V process, which is crucially important for the success of a MAS design and implementation. The presented attitudes are based on the experience described in the published literature and the authors' experience, which are[9]:

- ➢ V&V outcome should not be considered as a binary variable where the MAS are correct or incorrect
- ➢ acceptability and accreditation of MAS implementation results needs well-formulated problem .
- ➢ According to predefined objectives a MAS is built and to those objectives its credibility is judged .
- ➢ Throughout the entire MAS life cycle V&V must be conducted.
- ➢ To prevent developer's bias V&V require independence.
- ➢ V&V must be planned and documented.
- ➢ Successfully testing each agent does not imply credibility of the system.
- ➢ in MAS life cycle Errors should be detected as early as possible.

### 2.3. V&V constituents

The V&V constituent are [5]:

- *Correctness:* The system should be 100% correct.
  different human experts may have vary opinions on systems correctness .
- *Usability:* The system should meet user's demand.
- *Reliability:* How often the system fails to arrive at the correct problem's solution.
- *Consistency:* The requirement specification or system is free of internal contradiction
- *Competency:* The quality of the knowledge in a system relative to human skills.
- *Completeness:* Is a measure of the portion of specification implemented in the system.
- *Testability:* The system must designed in such a way to permit a testing plan to be carried out.
- *Adaptability:* How closely the system tied to a single model of work.

## 3. V&V IN THE MAS LIFE CYCLE

V&V is not a step or phase in the MAS life cycle, but it is a continuous activity throughout the entire life cycle as stated in attitudes above. (Figure 3) illustrate V&V activities in the MAS life cycle.

The MAS life cycle should not be interpreted as strictly sequential. sequential representation of some arrows is intended to show the direction of life cycle development , which is iterative in nature and reverse transitions are expected. Conducting V&V for the first time in the life cycle when the MAS is completed eliminates the opportunities to notify the system insufficiencies. Severe problems in the life cycle may undiscovered early until it become too late to do anything to fix and solve them.

Frequent tests throughout the system lifecycle are intended to inform the developers about their insufficiencies. The V&V activities throughout the entire MAS life cycle are intended to reveal any quality insufficiencies that might be present as the MAS progresses from the problem definition to the completion of the MAS application. This allows us to identify quality insufficiencies during the life cycle phase in which they occur, as stated in rules above.

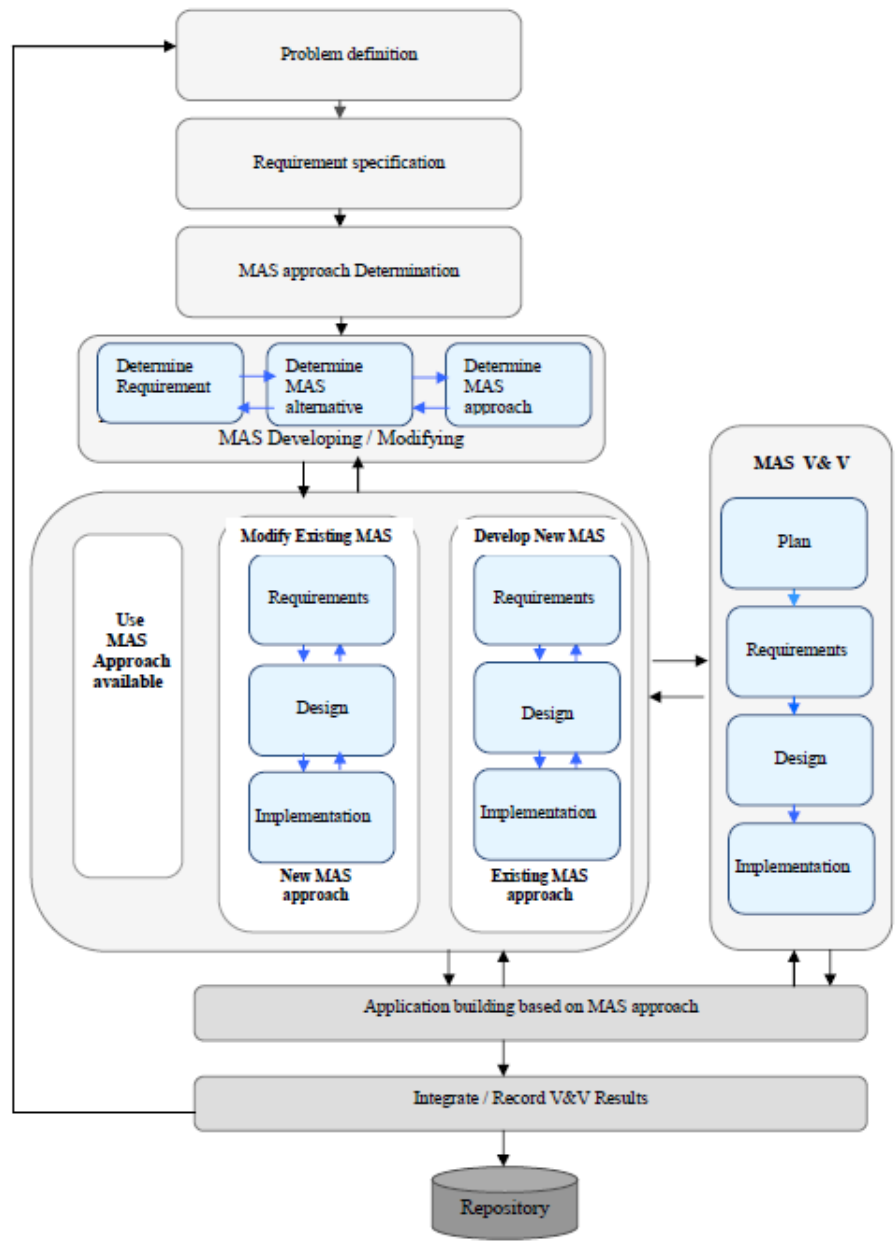

Figure 3. V&V activities in the MAS life cycle

## 3.1. V&V plan

Determine the V&V plan is the first step in the MAS V&V, in this plan describes the V&V activities that will take place during the MAS life cycle.

## 3.2. MAS Requirements V&V

The MAS requirements V&V activity is conducted to ensure that the specified requirements are accurate, readable, testable ,correct, complete, consistent, and satisfy the MAS requirements.

119

Poorly specified requirements (e.g., incorrect, ambiguous, incomplete, or not testable) contribute costly system cost overruns and unreliable. documentation of Inputs to the MAS requirements V&V activity are written in natural or formal mathematical languages and may be included graphics and charts. the complexity of the problems being solved make identifying MAS requirements difficult because causes uncertainty in developing the intended system performance requirements.

Repetition changes in requirements (e.g. change in environment to incorporate, new technologies, new tasks, new knowledge, new interfacing systems) throughout the MAS development process adds significantly more chance for error. Avoiding these problems, MAS requirements V&V verify the early system documentation (e.g., the System Requirements Specification) for feasibility and will use the rules, conventions, algorithms, and practices appropriate to the system application domain .

The MAS requirements V&V activity depends on the following activities:

- Compliance: Evaluate software requirements for to software requirements standards and software engineering practices.
- Satisfies user needs: Evaluate the defined concept to determine whether it satisfies user needs and system objectives in terms of system performance requirements, feasibility, completeness, and accuracy.
- Constraints: for interfacing systems identify major constraints and for the proposed approach identify constraints/limitations .
- Traceability analysis: Conduct a system traceability analysis by tracing the system requirements to system objectives and vice versa.
- Requirements evaluation: Conduct a system requirements evaluation by measuring the completeness and the correctness of the requirements.
- Requirements criticality: critical areas of software are Identify by assessing software requirements criticality.
- Interface analysis: Conduct a system interface analysis to evaluate the system requirements with respect to hardware, user and software interface requirements for accuracy, completeness, consistency, correctness, and Understandability.
- Domain Verifying: verify the scope and complexity of the proposed domain for the system knowledge base by verifying the correctness of the requirements on accuracy and completeness of the expected results.
  It is worth noting that, the MAS design V&V activity is modified if the developer chooses to reuse existing MAS requirements to include the following activities:
- Conduct an evaluation of the original MAS requirement documentation for compliance to MAS requirements of the new system.
- Conduct system interface analysis to evaluate reused system to new requirements for accuracy, completeness, consistency, correctness, and understandability.
- Generate the new system requirements or justify the use of the system without the required information.
- Evaluate whether or not the system requirements is adequate to support the new requirements.

### 3.3. MAS Design V&V

verify and validate that the MAS design meets the system requirements and will not affect performance of the system and it provides assurance that the system requirements are designed well and that other constraints are managed correctly.

International Journal of Software Engineering & Applications (IJSEA), Vol.3, No.5, September 2012

when the functional requirements and implementation constraints relating to data structures, accuracy , memory space, and timing are misrepresentation these can introduced design errors . To meet this requirement.

The following activities are implemented to ensure that the system requirements are not misrepresented, incompletely implemented, or incorrectly implemented:

  ☐ complete: Verify that the domain model is complete, consistent and represents the domain knowledge at the required level of accuracy and completeness.
  ☐ Traceability: Conduct a design traceability analysis to trace the MAS design to its requirements and vice versa and check the relationships for completeness, consistency, and correctness.
  ☐ Evaluation :Conduct a design evaluation to evaluate the MAS design for accuracy, completeness, consistency, correctness, and testability.
  ☐ the MAS design for compliance must evaluated with software engineering practices software design standards, and language standards.
  ☐ Interface analysis: Conduct a design interface analysis to evaluate the accuracy, completeness, consistency, and correctness

### 3.4. MAS Implementation V&V

It is performed to ensure that the MAS model defined by the system requirements and formalized during the system design is implemented in a complete, consistent and accurate way.  It involves code V&V activity, which is performed to verify and validate the correct implementation of the MAS design into code. It helps  finding and removing defects that may be cause unnecessary delays and costs from advancing poor code into any of the test activities.

The following activities must performed :

☐ Traceability analysis : source code traced  to software design, and vice versa.
☐  Evaluate  standardization :evaluate   source  code  for  compliance  with  code  standards, Language standards and software engineering practices.
☐ Interface analysis : evaluate the source code for accuracy, completeness, consistency, and correctness with respect to the hardware, user, and software interfaces
☐ Peer reviews: involve that the system in question is fulfilling its objectives, During a peer review, the reviewer's goal to identify any code that contains a potential defect And ensure that components implement their intended functionality using interfaces and methods specified in design.

### 3.5.  MAS Application V&V

Designed and implemented  the verification and validation of the application with respect to the specified MAS model. Examine the applicability of the MAS models through various structures, functions ,sizes, and application domains in different development and operational environments. At this point  the changes that may occur to the specified MAS requirements or design will lead to some V&V tasks. One of the key challenges in the MAS application V&V is the knowledge consistency.

### 4.6. Integrate and Record the V&V Results

the MAS V&V results are integrated and archived in repository . The changes that may occur to the specified MAS requirements or design or any possible changes needed in the system implementation will lead to reverse transitions back to the early stages.

## 5. V&V TECHNIQUES

| V&V techniques | Features | Example | Discussion |
|---|---|---|---|
| Formal | Represent information with Form formalisms, Non-ambiguous language Are based on mathematic Principles, like Z model, B method.[10][11] | -IMPRESS [12] that shows how to integrate different techniques into a common formal base. it originates in solving several problems that affects the validation tests of specifications | *Advantages:* - it allow drawing more conclusions from specifications in an automated way -proving complex Properties of the system and the domain, or the execution and animation specification *drawbacks:* - the specification needs a high level of detail in order to be used for the V&V techniques. - formal notations are difficult to get end users to validate and approve logic-based representat |
| Semiformal | - use some kind of semiformal notation, diagrams, to describe the information about their artifacts. -notations have syntaxes with a number of predefined primitives, and in a limited manner allow the use of natural language - vocabulary usually come With processes to elicit and transform the information describing it with the target notation. - resources about that information are quite limited, and they are just manual processes . | INGENIAS [13][14] is an agent-based system, that implements several meta-models which give the primitives and syntactic properties of models. it a methodology with solutions to verify properties. | *Advantages:* -It give better guidance for processes and more automated techniques. *drawbacks:* -advanced verifications of multiple properties are difficult because they imply a manual combination of checks where mistakes can be hard to detect. |
| Hybrid | -mix some of the semiformal and the formal techniques at the | *Tropos*[15], is agent based software methodology, *Tropos* use diagrams, | *Advantages:* -maintaining the usability of the semi-formal techniques |
| | same time.<br>- communication tool are used from semiformal techniques, are , while a fine-grained specification sophisticated computation over models is needed from the formal techniques | partially formal notations, and formal specifications.[16] | with a level of analysis close to formal when required<br>*drawbacks:*<br>-the nature of the properties to be checked,<br>as they come from the client requirements or the knowledge and experience of the developers<br>- these properties influence the specifications to make them consistent with software engineering practices and the needs identified by the client. |
| Conventional | -specifications in natural language complemented with arbitrary diagrams or images.<br>- Processes are described natural language.<br>-The accuracy of depends on the skills of the developer.[17][18] | Ethnography is used to verify and / or validate a set of design decisions using the real context in which they were taken.<br>- Ethnography depends on knowledge gathering through the observation of the system actors i the document. | *Advantages:*<br>- easy to use due to depend on the language' common symbols of the everyday communication.<br>*drawbacks:*<br>- results of its studies are not presented in a format suitable of reuse in a development as they do not directly address the developers' needs and do not describe clearly design issues.<br>- decisions about the process to follow belong to the personal criteria and are taken without the help of structured guidelines.<br>- the use of these languages does not guarantee a correct understanding in the development phases as the considered abstractions can be unknown nor have a different meaning for some developers. |

## 6. CONCLUSIONS

- Conducting the life cycle application of V&V is highly important for successful completion of large-scale and complex MAS. Applying the V&V techniques throughout the life cycle is time consuming and costly, therefore, continues research is needed to bring automation to the application of V&V techniques that must be clearly understood by the developer of the MAS.

- applying the MAS V&V approach and the implemented V&V techniques during the MAS life cycle are specified according to the objectives, problem domain, MAS architectural design model .

- Testing should continue until achieving sufficient confidence in credibility and acceptability of the MAS V&V results as dictated by the MAS objectives.

Our researches future work are to takes into account some features of agents like their motivation, agent learning capabilities, agents for different organization and choose the V&V technique during MAS V&V.

## ACKNOWLEDGMENTS

The authors would like to thank the Al-Zaytoonah Private University of Jordan in Amman, Jordan for Supporting this publication.

## REFERENCES

[1]   Dolor R.Wallace ,Laura M.Ippolito ,Barbara Cuthill(1996) "Reference Information for the Software Verification and Validation Process" NIST Special Publication
[2]   Wooldridge, M., Ciancarini, (2002) " Agent-Oriented Software Engineering" The State of the Art. In Ciancarini, P. and Wooldridge, M., editors, 1st International Workshop on Agent-Oriented Software Engineering, volume 1957,Springer-Verlag, Berlin.
[3]   M. Woolridge,( 2009)  "An Introduction to Multi-Agent-Systems,2nd edition", Wiley
[4]   Gerald D. Everet,Raymond McLeod, Jr.(2007) "Testing Across the Entire Software Development Life Cycle"
[5]   P. Ciancarini, M.J. Wooldridge,(2001)  "Agent-Oriented Software Engineering", Springer-Verlag.
[6]   Rubén Fuentes, Jorge J. Gómez-Sanz, Juan Pavón, (2002) "Verification and Validation Techniques for Multi-Agent Systems".
[7]   Wooldridge, M., Ciancarini, (2002) " Agent-Oriented Software Engineering" The State of the Art. 1st International Workshop on Agent-Oriented Software Engineering, volume 1957,Springer-Verlag, Berlin.
[8]   Paul Ammann, Jeff Offutt (2008) " Introduction  to Software Testing" .
[9]   I. Sommerville, (2011) " Software Engineering", 9th edition, Addison-Wesley.
[10]  Brazier, Dunin-Keplicz, et al. (1995), "Formal Specification of Multi-Agent Systems: A Real World Case, ICMAS"
[11]  H. Fadil and J. Koning,( 2005), " A formal approach to model multi agent interactions using the B formal method".  Fourth IEEE International Symposium on Advanced Distributed Systems, volume 3563 of LNCS.
[12]  M.M. West, T.L. McCluskey,(2001) " The application of machine learning tools to the validation of an air traffic control domain theory", International Journal on Artificial Intelligence Tools 10 .
[13]  J.G. Sanz, J. Pavón, (2003) ,"Agent oriented software engineering with INGENIAS", in: Proceedings of the CEEMAS.
[14]  Gomez-Sanz, J., Pavón J.(2003) "Agent Oriented Software Engineering with INGENIAS". In Proceedings of the CEEMAS
[15]  Ariel Fuxman, Lin Li , John Mylopoulos, Marco Pistore , Marco Roveri, Paolo Traverso ,(2004) "Specifying and Analyzing Early Requirements in Tropos"ITC-irst, Via Sommarive 18, I-38050, Trento, Italy.
[16]  Galvao Lourenco da Silva, I. (2005) "Design and Implementation of Multi-Agent Systems: The Tropos Case". Masters Thesis.
[17]  Rubén Fuentes, Jorge J. Gómez-Sanz, Juan Pavón ,(2004),"Verification and Validation Techniques for Multi-Agent Systems" The European Journal for the Informatics Professional , Volume 2004, No.4.
[18]  L.A. Carreira , S.H. Hilal and S.W. Karickhoff,(2003) "verification and validation of the SPARC model"